\documentclass[]{pasj01}

\begin{document} 
\Received{}
\Accepted{}

\title{High Precision Analyses of Ly$\alpha$ Damping Wing
of Gamma-Ray Bursts in the Reionization Era: On the Controversial
Results from GRB 130606A at $z = 5.91$}

\author{Tomonori \textsc{Totani},\altaffilmark{1}}
\altaffiltext{1}{Department of Astronomy, The University of Tokyo, 
     Hongo, Tokyo 113-0033}

\author{Kentaro \textsc{Aoki},\altaffilmark{2}}
\altaffiltext{2}{Subaru Telescope, National Astronomical Observatory of 
   Japan, 650 North A'ohoku Place, Hilo, HI 96720, USA}

\author{Takashi \textsc{Hattori},\altaffilmark{2}}

\author{Nobuyuki \textsc{Kawai}\altaffilmark{3}}
\altaffiltext{3}{Department of Physics, Tokyo Institute of Technology, 
    2-12-1 Ookayama, Meguro-ku, Tokyo 152-8551}

\KeyWords{Techniques: spectroscopic --- Gamma-ray burst: individual: 
GRB 130606A --- dark ages, reionization, first stars } 

\maketitle

\begin{abstract}
The unprecedentedly bright afterglow of Swift GRB 130606A at $z =
5.91$ gave us a unique opportunity to probe the reionization era by
high precision analyses of the redward damping wing of Ly$\alpha$
absorption, but the reported constraints on the neutral hydrogen
fraction ($f_{\rm H\emissiontype{I}}$) in intergalactic medium (IGM)
derived from spectra taken by different telescopes are in
contradiction. Here we examine the origin of this discrepancy by
analyzing the spectrum taken by VLT with our own analysis code
previously used to fit the Subaru spectrum.  Though the VLT team
reported no evidence for IGM H\emissiontype{I} using the VLT spectrum,
we confirmed our previous result of preferring non-zero IGM
H\emissiontype{I} (the best-fit $f_{\rm H\emissiontype{I}} \sim 0.06$,
when IGM H\emissiontype{I} extends to the GRB redshift).  The fit
residuals of the VLT spectrum by the model without IGM
H\emissiontype{I} show the same systematic trend as the Subaru
spectrum.  We consider that the likely origin of the discrepancy
between the two teams is the difference of the wavelength ranges
adopted in the fittings; our wavelength range is wider than that of
the VLT team, and also we avoided the shortest wavelength range of
deep Ly$\alpha$ absorption ($\lambda_{\rm obs} < 8426$ {\AA}), because
this region is dominated by H\emissiontype{I} in the host galaxy and
the systematic uncertainty about host H\emissiontype{I} velocity
distribution is large.  We also study the sensitivity of these results
to the adopted Ly$\alpha$ cross section formulae, ranging from the
classical Lorentzian function to the most recent one taking into
account fully quantum mechanical scattering.  It is found that the
preference for non-zero IGM H\emissiontype{I} is robust against the
choice of the cross section formulae, but it is quantitatively not
negligible and hence one should be careful in future analyses.
\end{abstract}

\section{Introduction}

Hydrogen gas in the intergalactic medium (IGM) was reionized at
redshifts $z \gtrsim 6$, and it is important to reveal when and how
this process occurred quantitatively for better understanding of the
formation and evolution of the earliest galaxy populations (Barkana \&
Loeb 2007; Meiksin 2009; Robertson et al. 2010; Fan 2012 for reviews).
Gamma-ray bursts (GRBs) are an important probe for the distant
universe reaching the epoch of cosmic reionization ($z \gtrsim 6$),
giving us useful information such as the cosmic star formation rate
(see Piran 2005; M\'esz\'aros 2006; Zhang 2007; Gehrels et al. 2009
for reviews).  When a high precision optical/near-infrared spectrum is
taken for a GRB afterglow in the reionization era, light at
wavelengths shorter than the redshifted Ly$\alpha$ is almost
completely absorbed by neutral hydrogen in IGM, and so-called
Gunn-Peterson troughs appear. Because of the strength of Ly$\alpha$
absorption, GP troughs give only a weak lower bound for the IGM
neutral fraction as $f_{\rm H\emissiontype{I}} \equiv n_{\rm
  H\emissiontype{I}}/n_{\rm H} \gtrsim 10^{-3}$. However, the shape of
redward Ly$\alpha$ damping wing can be used to further constrain
$f_{\rm H\emissiontype{I}}$ and hence the reionization history,
because IGM H\emissiontype{I} would affect the shape of the wing, in
addition to the wing by H\emissiontype{I} in the host galaxy
(Miralda-Escude 1998).  GRBs have a few advantages against quasars
about this test.  They are expected to occur in more normal and less
dense regions than quasars, and short duration of GRB emission does
not affect the ionization status around host galaxies. The simple
power-law spectrum of GRB afterglows is suitable for a precise fit to
the damping wing shape.

However, use of GRBs for reionization study has been hampered by the
low event rate of high-$z$ GRBs that are bright enough for high
precision damping wing analyses. A weak upper bound on IGM
H\emissiontype{I} was obtained for GRB 050904 at $z = 6.3$ (Kawai et
al. 2006; Totani et al. 2006), but meaningful constraints on
reionization could not be derived by GRBs at even higher redshifts
(Greiner et al. 2009; Patel et al. 2010; Salvaterra et al. 2009;
Tanvir et al. 2009; Cucchiara et al. 2011) because of their low
signal-to-noise ratio. The discovery of GRB 130606A (Ukwatta et
al. 2013; Golenetskii et al. 2013; Castro-Tirado et al. 2013) at $z =
5.91$ provided us with the best opportunity so far for this purpose,
by its exceptional brightness and relatively low H\emissiontype{I}
column density in the host galaxy\footnote{After this event, GRB
  140515A was detected at $z=6.33$ and constraints on reionization
  have been derived (Chornock et al. 2014; Melandri et al. 2015), but
  the afterglow was not as bright as that of GRB 130606A.}.  However,
the reported constraints on $f_{\rm H\emissiontype{I}}$ using spectra
taken by different telescopes are controversial.  Chornock et
al. (2013) derived an upper bound of $f_{\rm H\emissiontype{I}} <
0.11$ ($2\sigma$) using the spectrum taken by Gemini. In contrast,
Totani et al. (2014, hereafter Paper I) found a $\sim 3 \sigma$
evidence for IGM H\emissiontype{I} with $f_{\rm H\emissiontype{I}} 
\gtrsim 0.08$ from the spectrum
taken by Subaru. It was argued that the choice of the baseline
power-law index $\beta = -1.99$ ($f_\nu \propto \nu^\beta$) in
Chornock et al. is not supported by the observed optical/NIR colors
favoring $\beta \sim -1$. This has been confirmed by the optical-NIR
spectrum by VLT, reporting $\beta = -1.02 \pm 0.03$ (Hartoog et
al. 2014, hereafter H14).  

However, this is not the end of the story. The analysis of the damping
wing by H14 using the VLT spectrum found no evidence for IGM
H\emissiontype{I}, setting an upper limit on $f_{\rm
  H\emissiontype{I}} < 0.03$ ($3 \sigma$).  This is statistically
inconsistent with the result of our Paper I, indicating the systematic
uncertainties in the Subaru/VLT spectra and/or analysis methods. GRB
130606A has proven that a high precision damping wing analysis for GRB
spectra in the reionization era is indeed possible, and understanding
systematics in such analyses is crucial for future studies using more
GRBs at higher redshifts to derive reliable constraints on
reionization. Therefore we decided to analyze the VLT spectrum by our
own code used for the fitting to the Subaru spectrum in Paper I, to
reveal the origin of the discrepant results. This is the primary aim
of this work.

Another aim of this work is to examine the effect of adopted formulae
for the Ly$\alpha$ cross section as a function of wavelength. Several
different formulae have been used in the literature to calculate the
damping wing shape by H\emissiontype{I} in a host galaxy and IGM,
including the simplest Lorentzian (or the Voigt profile when convolved
with the Gaussian velocity distribution), and the two-level
approximation formula by Peebles (1993, hereafter P93).  In high
precision damping wing analyses, difference of the adopted formulae
may result in systematic biases. Here we repeat our analyses using
different cross section formulae, including the most recent formula by
Bach \& Lee (2015, hereafter BL15) that takes into account the fully
quantum mechanical scattering based on the second-order time-dependent
perturbation theory, and see how the results change.

Unless otherwise stated, we adopt the same model parameter values as
those in Paper I. The H\emissiontype{I} column density in the host galaxy,
$N_{\rm H\emissiontype{I}}^{\rm host}$, is expressed in units of cm$^{-2}$.

\section{On the Controversial Results between Subaru and VLT}

\subsection{Comparison between the Subaru and VLT Spectra}

\begin{figure}
 \begin{center}
  \includegraphics[width=13cm,angle=-90]{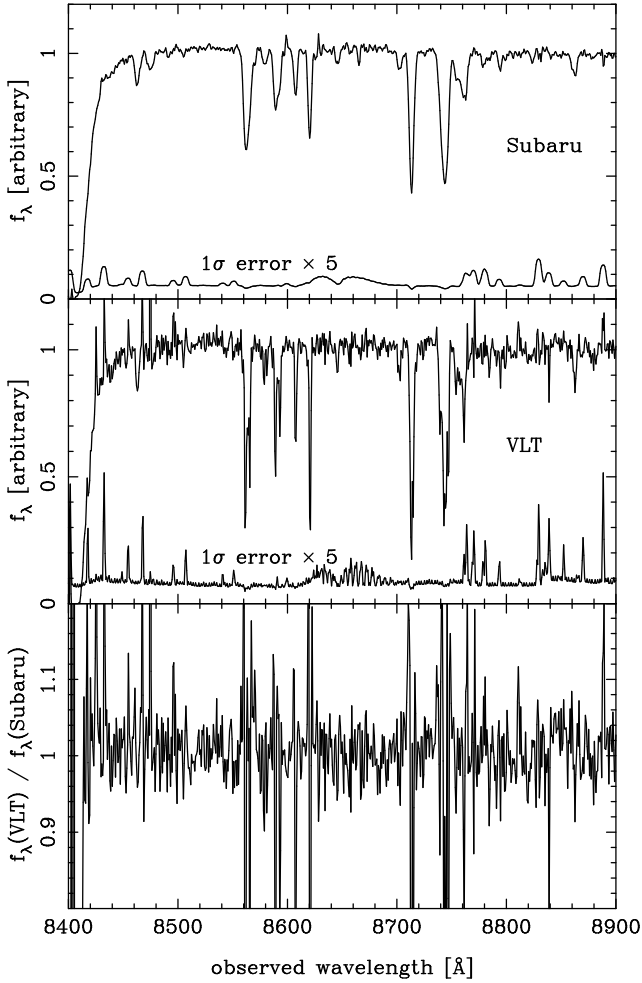} 
 \end{center}
\caption{Top: the Subaru spectrum of the afterglow of GRB 130606A
  reported in Paper I. The 1$\sigma$ error is shown, but multiplied by
  a factor of five for presentation purpose.  Middle: the same as the
  top panel but for the VLT spectrum of H14.  Here, the VLT spectrum
  has been converted on the wavelength grids of the Subaru spectrum
  (see text).  Note that the spectral resolution of the VLT spectrum
  is better than the Subaru even though the wavelength grids are the
  same.  Bottom: the flux ratio of the VLT to the Subaru spectrum.}
\label{fig:vlt_vs_subaru}
\end{figure}

First we directly compare the Subaru and VLT spectra of GRB 130606A,
which were taken during 10.4--13.2 and 7.2--8.7 hr after the burst,
respectively. We use the one-dimensional VLT spectrum and its
statistical error, which were reduced by the VLT team and provided to
us.  The original VLT spectrum has a wavelength binning size of 0.20
{\AA}, which is smaller than 0.74 {\AA} for the Subaru spectrum.  To
compare the two spectra on the same wavelength grids, the VLT spectrum
was converted into that on the Subaru grids. This was done first by
associating every VLT grid to the closest Subaru grid, and then by
taking the average over the VLT grids associated to a Subaru grid. The
two spectra on the same grids are then shown in
Fig. \ref{fig:vlt_vs_subaru}.  The $1\sigma$ error at regions without
strong airglow emission is typically 1.0\% and 1.5\% of the continuum
level for Subaru and VLT, respectively, on this same wavelength grids.
The VLT spectrum shows finer structure because of the better spectral
resolution ($R = \lambda/\Delta \lambda \sim 2200$ and 8700,
respectively).  The ratio of the two spectra is also shown in this
figure.  Though there are fluctuations on $\sim$ 10 {\AA} scale, no
systematic trend is found on scales longer than 100 {\AA}.

\subsection{Results of the Damping Wing Fits}

In Paper I, we considered two different models for the IGM
H\emissiontype{I} distribution; in one model, the IGM
H\emissiontype{I} is assumed to extend up to the same redshift as the
GRB, i.e., $z_{\rm IGM, u} = z_{\rm GRB} = 5.9131$, while in the other
model, $z_{\rm IGM, u}$ is set at 5.83 motivated by a deep GP trough
observed at $z = $ 5.67--5.83 to the sightline of this GRB (Chornock
et al. 2013).\footnote{The fit is insensitive to the redshift lower
  bound $z_{\rm IGM, l}$ if $z_{\rm IGM, l} \ll z_{\rm IGM, u}$, and
  contribution to the damping wing is mostly from a redshift interval
  of $\Delta z \sim 0.1$ from $z_{\rm IGM, u}$ [see Fig. 7 of Totani
    et al. (2006)], corresponding to a proper distance of $\sim 6$
  Mpc.}  It should be noted here that the damping wing shape is
calculated assuming that IGM H\emissiontype{I} is uniformly
distributed with a constant comoving density, which is obviously too
simple compared with the expected realistic distribution.  Though it
is not the purpose of this paper to examine this effect, this should
be kept in mind when the implications for reionization are
discussed. The best-fit model parameters are $(f_{\rm
  H\emissiontype{I}}, \beta) = (0.086^{+0.012}_{-0.011},
-0.93^{+0.04}_{-0.04})$ and $(0.47^{+0.08}_{-0.07},
-0.74^{+0.09}_{-0.07})$ for the former and latter, respectively.  The
$\chi^2$ difference between the two models is not statistically
significant, but both the two models show a $\sim 3 \sigma$
statistical preference to the model without IGM H\emissiontype{I}. It
should be noted that some other possibilities of reddening absorption
[extinction in the host galaxy or intervening damped Ly$\alpha$
  systems (DLAs)] have been ruled out in Paper I.

\begin{figure*}
 \begin{center}
  \includegraphics[width=13cm,angle=-90]{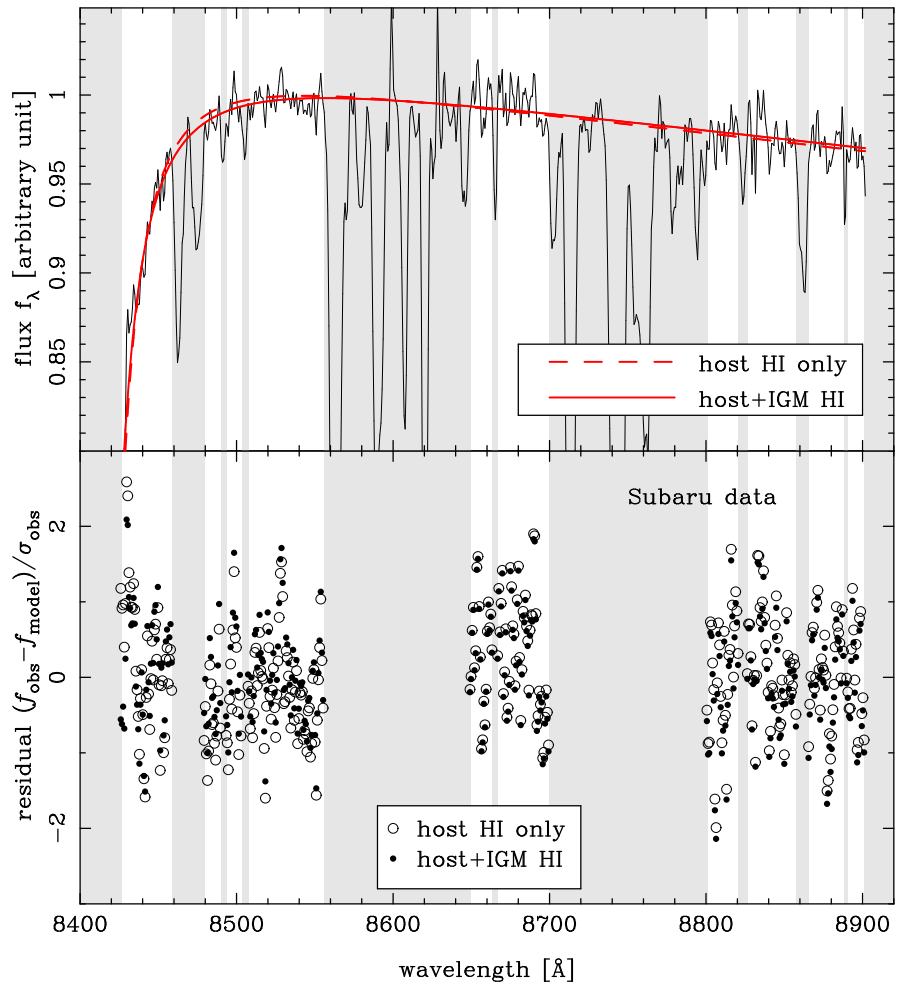} 
 \end{center}
\caption{Top: the observed spectrum taken by Subaru is shown in thin
  solid line. The best-fit curves of the host H\emissiontype{I} only
  model and the host+IGM H\emissiontype{I} model are shown by thick
  dashed and solid curves, respectively.  Bottom: the fit residuals
  $(f_{\rm obs} - f_{\rm model})/\sigma_{\rm obs}$ are shown for the
  two models.  The gray shaded regions indicate the wavelength ranges
  removed from the fits because of apparent features of absorption
  lines or airglow.  }
\label{fig:fit_subaru}
\end{figure*}

\begin{figure*}
 \begin{center}
  \includegraphics[width=13cm,angle=-90]{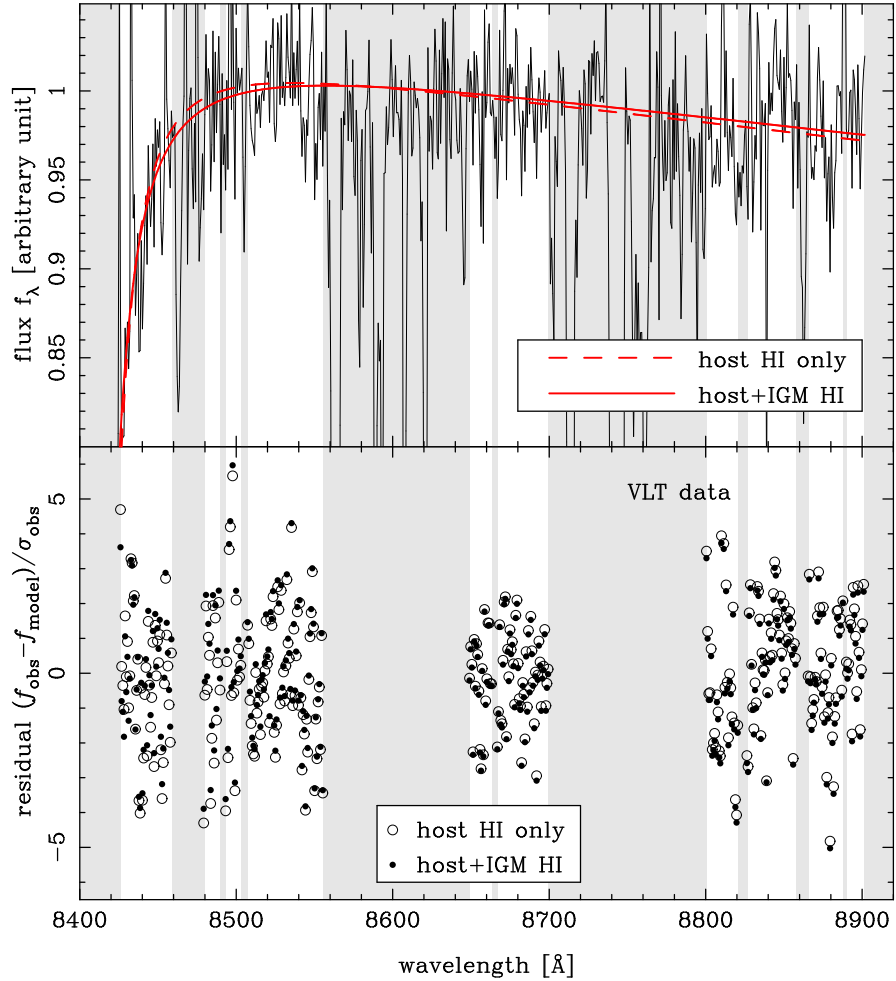} 
 \end{center}
\caption{The same as Fig. \ref{fig:fit_subaru}, but
for the VLT spectrum converted onto the wavelength grids
of the Subaru spectrum.
}
\label{fig:fit_vlt}
\end{figure*}

However, H14 reported $\beta = -1.02 \pm 0.03$ from their optical+NIR
spectrum, indicating that the low $z_{\rm IGM,u}$ model in Paper I is
now disfavored.  Therefore in this paper we consider only the $z_{\rm
  IGM, u} = z_{\rm GRB}$ model with the power index fixed at $\beta =
-1.02$, to compare the Subaru and VLT spectra\footnote{The error of
  $\beta$ reported by H14 is smaller than the typical statistical
  uncertainties of free $\beta$ fits in Paper I, and hence we do not
  marginalize the error of the H14 $\beta$ measurement.}.  Then the
model parameters are: the column density of H\emissiontype{I} in the
host galaxy $N_{\rm H\emissiontype{I}}^{\rm host}$, the 1$\sigma$
Gaussian velocity dispersion $\sigma_v$ of H\emissiontype{I} in the
host galaxy, $f_{\rm H\emissiontype{I}}$ of IGM H\emissiontype{I}, and
the overall normalization factor.  The fit results of the host
H\emissiontype{I} only model ($f_{\rm H\emissiontype{I}}$ fixed to
zero) and the host+IGM H\emissiontype{I} model ($f_{\rm
  H\emissiontype{I}}$ treated as a free parameter) to the Subaru and
VLT spectra are shown in Table \ref{table:subaru_vs_vlt} and Figs
\ref{fig:fit_subaru} and \ref{fig:fit_vlt}.  Here, we used the VLT
spectrum converted onto the wavelength grids of the Subaru spectrum as
described in the previous section, and exactly the same analysis
procedures as Paper I were adopted to the two spectra.

\begin{table*}
\tbl{The best fit parameters of the fittings to the Subaru
and VLT spectra$^*$}{
\begin{tabular}{lccccc}
\hline
\hline
model  & $\log_{10} (N_{\rm H\emissiontype{I}}^{\rm host})^\dagger$ &
 $\sigma_v$ (km/s)$^\ddagger$ & IGM $f_{\rm H\emissiontype{I}}$ & ${\chi^2}^\S$ 
 & $\Delta {\chi^2}^\|$ \\
\hline
\multicolumn{6}{c}{fit to the Subaru spectrum} \\
\hline
host H\emissiontype{I} only  
  &  19.877$^{+0.008}_{-0.015}$ & 0.0$^{+89.9}_{-0.0}$ 
  & fixed to zero & 95.10 & 14.48 \\
host+IGM H\emissiontype{I}
  &  19.768$^{+0.032}_{-0.032}$ 
  & 62.0$^{+38.0}_{-62.0}$ & 0.061$^{+0.007}_{-0.007}$ & 80.62 & - \\
\hline
\multicolumn{6}{c}{fit to the VLT spectrum} \\
\hline
host H\emissiontype{I} \ only
  &  19.806$^{+0.014}_{-0.016}$ 
  & 0.0$^{+52.0}_{-0.0}$  & fixed to zero & 292.57 & 11.89 \\
host+IGM H\emissiontype{I}
  &  19.621$^{+0.059}_{-0.057}$ 
  & 0.0$^{+100.0}_{-0.0}$ & 0.087$^{+0.017}_{-0.029}$ & 280.68 & - \\
\hline
\hline
\end{tabular} }
\label{table:subaru_vs_vlt}
\begin{tabnote}
$^*$The fit results for the Ly$\alpha$ damping wing models including
  only H\emissiontype{I} in the host galaxy (``host H\emissiontype{I}
  only'') and including host plus IGM H\emissiontype{I} (``host+IGM
  H\emissiontype{I}'', with IGM H\emissiontype{I} extending up to the
  redshift of the GRB). The quoted
  errors are statistical 1$\sigma$. \\ $^\dagger$The neutral hydrogen
  column density in the GRB host ($N_{\rm H\emissiontype{I}}^{\rm
    host}$) in units of cm$^{-2}$. \\ $^\ddagger$The survey range of
  $\sigma_v$ ($1 \sigma$ of the Gaussian velocity distribution of host
  H\emissiontype{I}) is limited to 0--100 km/s, motivated from the
  observed widths of the metal absorption lines (Paper I).
  \\ $^\S$The number of data points is 68.
  \\ $^\|$The $\chi^2$ excess of the host H\emissiontype{I} only model
  against the host+IGM H\emissiontype{I} model.
\end{tabnote}
\end{table*}

The fit results to the Subaru spectrum with the fixed value of $\beta
= -1.02$ are similar to the $z_{\rm IGM, u} = z_{\rm GRB}$ model with
free $\beta$ reported in Paper I; a non-zero IGM contribution with
$f_{\rm H\emissiontype{I}} \sim $ 0.06 is favored than the host
H\emissiontype{I} only model.  The relative flux difference between
the two models is only $\sim$ 0.5\% level, but the overall statistical
significance is $(\Delta \chi^2)^{1/2} = 3.8 \sigma$ thanks to the
small statistical error of the spectrum. It should be noted that the
errors on the Subaru spectrum are uncorrelated among different
wavelength bins and the scatter is consistent with the Gaussian (Paper
I). There are 68 data points in the fit, and there are four model
parameters in the host+IGM H\emissiontype{I} model.  The $\chi^2$
value of 80.62 for $68-4 = 64$ degrees of freedom is not unreasonable
(8\% chance probability of getting a larger value).

The reason why the IGM H\emissiontype{I} component improves the fit
can be seen in the residuals (normalized by the observed flux error)
shown in Fig. \ref{fig:fit_subaru}.  The fit residuals of the host
only model tend to be positive at longer wavelength range of
8800--8900 \AA, but tend to be negative at 8480--8560 \AA, indicating
that another absorption component is necessary to make the spectrum
further redder.  It is impossible to do this by increasing the
amount of host H\emissiontype{I}, because it would result in too large
absorption in the shortest wavelength range of 8420--8460 \AA. Since
H\emissiontype{I} in IGM is more distant from the GRB than
H\emissiontype{I} in the host, the damping wing by IGM
H\emissiontype{I} has weaker wavelength dependence than that by host
H\emissiontype{I} (see Fig. 2 of Paper I). This is the key for the fit
improvement by IGM H\emissiontype{I}.  It should also be noted that a
reduction of the scatter of residuals by IGM H\emissiontype{I} is seen
in 8420--8460 \AA.  This indicates that the damping wing shape with
IGM H\emissiontype{I} is in better agreement with the observed data
than the host H\emissiontype{I} only model, giving a further support
for the model with IGM H\emissiontype{I}.

Now we examine the fit results to the VLT spectrum.  We confirmed the
same result as that to the Subaru spectrum; non-zero IGM
H\emissiontype{I} contribution is favored with the best fit value of
$f_{\rm H\emissiontype{I}} = 0.087^{+0.017}_{-0.029}$ which is
statistically consistent with $f_{\rm H\emissiontype{I}} =
0.061^{+0.007}_{-0.007}$ found with the Subaru spectrum.  The
residuals of the fit to the VLT spectrum in Fig. \ref{fig:fit_vlt}
show the same systematic trend as those of the Subaru fit (in
Fig. \ref{fig:fit_subaru}).  Therefore the evidence for IGM
H\emissiontype{I} contribution to the observed damping wing reported
in Paper I is further strengthened by an independent spectrum taken by
VLT.  The $\chi^2$ difference implies that the host+IGM
H\emissiontype{I} model is preferred to the host only model with a
statistical significance of $11.89^{1/2} = 3.4 \sigma$.  However, the
residuals shown in Fig. \ref{fig:fit_subaru} are on average
significantly larger that the expectation of the Gaussian
distribution, suggesting that the errors calculated by the VLT
analysis pipeline are underestimated.  We discussed about this with
the VLT team, and it is most likely because the lower signal-to-noise
ratio with respect to the Subaru spectrum makes the systematic noise
due to the bright sky emission lines more prominent.  The VLT spectrum
was taken with an exposure time considerably shorter than that for
Subaru, and the air mass was also larger.  The median of absolute
values of the residuals for the host+IGM H\emissiontype{I} model is
1.221, while 0.674 is expected for the Gaussian distribution.  If we
scale the error size by the ratio of 1.221/0.674, the significance is
reduced to 1.9$\sigma$. In this work we do not discuss the statistical
significance about the VLT spectrum further.

These results indicate that the discrepant conclusions (evidence for
IGM H\emissiontype{I} by Paper I, while no evidence found by H14) are
a result of difference in analysis methods, rather than the systematic
difference between the two spectra.  Concerning this, we note that the
wavelength range fitted by H14, 8406--8462 \AA \ (corresponding to
relative velocities of 0--2000 km/s to Ly$\alpha$), is significantly
shorter compared with 8426--8900 \AA \ adopted in our Paper I. It
should be noted that the evidence for IGM H\emissiontype{I} in our
analysis is found in the spectral shape spanning in the whole
wavelength range.  Another difference is that we excluded the shortest
wavelength range of $\lambda_{\rm obs} < $ 8426 \AA.  As shown in
Fig. 2 of Paper I, the H\emissiontype{I} absorption in this region is
dominated by the host H\emissiontype{I} rather than the IGM component
with $f_{\rm H\emissiontype{I}} \sim 0.1$.  The observed spectrum
rapidly drops with decreasing wavelength at $\lambda_{\rm obs} < 8426$
\AA, making a precise model fitting difficult because of uncertainties
about the velocity distribution of H\emissiontype{I} in the host and
spectral resolution.  Though the Gaussian velocity distribution has
been taken into account for the host H\emissiontype{I}, such a pure
Gaussian distribution is unlikely in a realistic galaxy, as inferred
from high resolution spectroscopic studies of DLAs (Wolfe et al. 2005;
see also Paper I for more discussion). This is why in Paper I we used
only wavelength longer than $8426$ \AA; this boundary was determined
so that the fit result becomes mostly insensitive to the host
H\emissiontype{I} velocity dispersion, $\sigma_v$ (see Table
\ref{table:subaru_vs_vlt}).

To test the effect of the wavelength range used, we performed the
damping wing fit to the Subaru spectrum with the same wavelength range
as H14.  Indeed, we found a similar result to H14; the best fit is
$f_{\rm H\emissiontype{I}} = 0$ with upper bounds of $f_{\rm
  H\emissiontype{I}} < 0.0034$ and 0.088 at 1 and 2$\sigma$,
respectively, $\log_{10} N_{\rm H\emissiontype{I}}^{\rm host}$ =
19.855$\pm$0.009, and $\sigma_v = 61.8 \pm 3.3$ km/s. The small error
of $\sigma_v$ indicates that the fit in this region is highly
sensitive to the host H\emissiontype{I} velocity distribution, but
systematic uncertainty should be large if the distribution is not a
simple Gaussian.

\section{Testing Different Ly$\alpha$ Cross Section Formulae}

In Paper I we used the two-level approximation formula of P93 for the
Ly$\alpha$ cross section as a function of wavelength.  In the
calculation of the damping wing by IGM H\emissiontype{I}, we further
used the analytic formula of Miralda-Escude (1998), which also assumes
the P93 formula and uses an approximation valid when the frequency is
far from the resonance $(4 \pi |\nu - \nu_\alpha| \gg \Gamma_\alpha)$
to make the analytic integration possible, where $\Gamma_\alpha =
6.262 \times 10^8 \ \rm s^{-1}$ is the damping constant of Ly$\alpha$.
Though this approximation is very good (corresponding to $|\lambda -
\lambda_\alpha|/\lambda_\alpha \gg 2.02 \times 10^{-8}$) for the
practical damping wing analysis, we revised our code to numerically
integrate the IGM damping wing, to calculate it for arbitrary formulae
of Ly$\alpha$ cross section.

\begin{figure}
 \begin{center}
  \includegraphics[width=13cm,angle=-90]{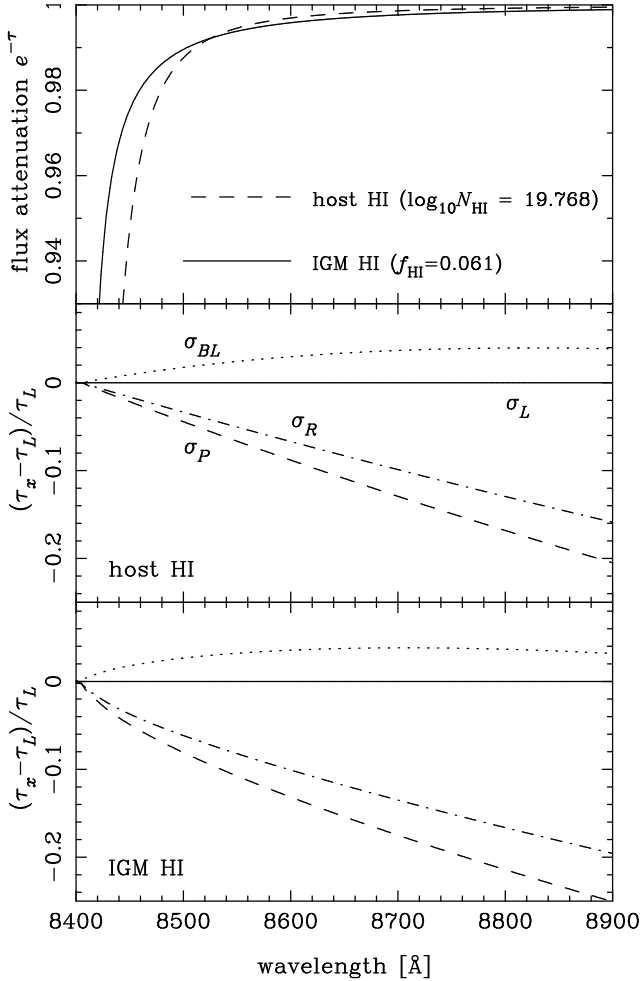} 
 \end{center}
\caption{Top: the flux attenuation factor $e^{-\tau}$ by the optical
  depth $\tau$ of absorption by H\emissiontype{I}, separately for the
  host galaxy H\emissiontype{I} and IGM H\emissiontype{I} with the
  best-fit parameters of the host+IGM H\emissiontype{I} model to the
  Subaru spectrum (Table \ref{table:subaru_vs_vlt}), using the
  Lorentzian cross section formula.  Middle: the fractional difference
  of $\tau$ by H\emissiontype{I} in the host galaxy when various
  Ly$\alpha$ cross section formulae ($\sigma_P$: Peebles, $\sigma_R$:
  Rayleigh, $\sigma_{BL}$: Bach-Lee) are used instead of the
  Lorentzian ($\sigma_L$).  Bottom: the same as the middle panel, but
  for $\tau$ by IGM H\emissiontype{I}. }
\label{fig:cross_section}
\end{figure}

Here we test three different formulae of Ly$\alpha$ cross section: the
Lorentzian, the classical Rayleigh scattering formula, and the fitting
formula of BL15 taking into account the quantum mechanical scattering
effect, in addition to the P93 formula used in Paper I.  (See BL15 for
the exact expressions of these formulae.)
Fig. \ref{fig:cross_section} shows the change of H\emissiontype{I}
optical depth as a function of wavelength for different formulae,
using the best-fit parameters of the host+IGM H\emissiontype{I} model
to the Subaru spectrum of GRB 130606A.  The change relative to the
Lorentzian becomes larger with increasing wavelength, reaching
10--20\% at 8900 \AA. The Rayleigh and P93 formulae result in smaller
optical depth, but the BL15 formula larger, compared with the
Lorentzian. Then we repeated our damping wing fitting analysis on the
Subaru spectrum for the host H\emissiontype{I} only and host+IGM
H\emissiontype{I} models, and $\chi^2$ are presented in Table
\ref{table:cross_section}.  The best-fit value of $f_{\rm
  H\emissiontype{I}}$ changes by less than 8\%.  The BL15 formula
results in the smallest significance of the statistical preference for
the non-zero $f_{\rm H\emissiontype{I}}$ to the host H\emissiontype{I}
only model, but still it is larger than $3.1 \sigma$.

\begin{table*}
\tbl{The best fit parameters in the fittings 
with different Ly$\alpha$ cross section formulae$^*$}{
\begin{tabular}{lccccc}
\hline
\hline
cross section formula  & $\log_{10} (N_{\rm H\emissiontype{I}}^{\rm host})$ &
 $\sigma_v$ (km/s) & IGM $f_{\rm H\emissiontype{I}}$ & $\chi^2$ 
 & $\Delta \chi^2$ \\
\hline
\multicolumn{6}{c}{host H\emissiontype{I} only model} \\
\hline
Lorentzian  
  &  19.869$^{+0.010}_{-0.010}$ & 0.0$^{+70.2}_{-0.0}$ 
  & fixed to zero & 91.81 & 10.74 \\
Rayleigh
  &  19.875$^{+0.010}_{-0.010}$ & 22.1$^{+63.1}_{-22.1}$ 
  & fixed to zero & 94.21 & 13.50 \\
Peebles  
  &  19.877$^{+0.008}_{-0.015}$ & 0.0$^{+89.9}_{-0.0}$ 
  & fixed to zero & 95.10 & 14.48 \\
Bach \& Lee  
  &  19.866$^{+0.009}_{-0.009}$ & 0.0$^{+63.5}_{-0.0}$ 
  & fixed to zero & 90.66 & 9.88 \\
\hline
\multicolumn{6}{c}{host + IGM H\emissiontype{I} model} \\
\hline
Lorentzian
  &  19.755$^{+0.033}_{-0.033}$ 
  & 100.0$^{+0.0}_{-100.0}$ & 0.057$^{+0.0012}_{-0.007}$ & 81.07 & - \\
Rayleigh
  &  19.765$^{+0.033}_{-0.033}$ 
  & 54.6$^{+45.4}_{-54.6}$ & 0.060$^{+0.008}_{-0.007}$ & 80.71 & - \\
Peebles
  &  19.768$^{+0.032}_{-0.032}$ 
  & 62.0$^{+38.0}_{-62.0}$ & 0.061$^{+0.007}_{-0.007}$ & 80.62 & - \\
Bach \& Lee
  &  19.751$^{+0.029}_{-0.029}$ 
  & 100.0$^{+0.0}_{-100.0}$ & 0.056$^{+0.011}_{-0.006}$ & 80.78 & - \\
\hline
\hline
\end{tabular} }
\label{table:cross_section}
\begin{tabnote}
$^*$The fitting results of the host H\emissiontype{I} only and
  host+IGM H\emissiontype{I} models for the Ly$\alpha$ damping wing,
  using the Subaru spectrum. See also Table \ref{table:subaru_vs_vlt}
  for more explanations.
\end{tabnote}
\end{table*}

\section{Discussion and Conclusions}

This paper investigated the origin of the discrepant results about the
constraint on IGM H\emissiontype{I} fraction from Ly$\alpha$ damping
wing analyses of the Subaru and VLT spectra for GRB 130606A at $z =
5.913$.  Direct comparison between the two spectra does not show any
systematic difference on the overall shape in the wavelength range of
8426--8900 \AA \ used in Paper I. We repeated exactly the same
analysis as Paper I for the VLT spectrum converted onto the
wavelength grids points of the Subaru spectrum, and confirmed our
previous result of more than $3 \sigma$ statistical preference for
non-zero contribution from IGM H\emissiontype{I} with $f_{\rm
  H\emissiontype{I}} \sim $0.05--0.1 (assuming uniform distribution of
IGM H\emissiontype{I} up to the redshift of the GRB).  The absorption by
IGM is dominated by H\emissiontype{I} within $\Delta z \sim 0.1$ from
the upper redshift bound ($z_{\rm IGM, u} = z_{\rm GRB}$),
corresponding to a proper distance of $\sim 6$ Mpc.  The host
H\emissiontype{I} only model is disfavored because of the systematic
trend of fit residuals (positive at 8640--8900 \AA, while negative at
8450--8560 \AA) indicating that another absorption component is
necessary to make the spectrum redder, in addition to
H\emissiontype{I} in the host galaxy. This trend has been confirmed
also using the VLT spectrum.

Therefore we consider that the discrepant results were obtained
because of different analyses methods. H14 adopted the wavelength
range of 8406--8462 \AA, which overlaps only with the shortest part of
the range adopted by us in Paper I, and H14 also included the deep
Ly$\alpha$ absorption region down to the resonant Ly$\alpha$
wavelength (8406 \AA). In Paper I we excluded the wavelengths shorter
than 8426 \AA \ because this region is highly sensitive to the
velocity distribution of H\emissiontype{I} in the host galaxy, which
is rather uncertain and unlikely to be a pure Gaussian. Indeed, we
found a result consistent with H14 (i.e., no evidence for non-zero
$f_{\rm H\emissiontype{I}}$) when we adopted the same wavelength range
as H14.

Finally, we examined the robustness of these results against the
adopted Ly$\alpha$ cross section formulae, by testing the Lorentzian,
Rayleigh scattering formula, and the latest BL15 fitting formula
taking into account a fully quantum mechanical scattering effect, in
addition to the Peebles' two-level approximation formula used in Paper
I.  We found that the preference for the non-zero $f_{\rm
  H\emissiontype{I}}$ is robust against the difference of these
formulae. The BL15 formula results in the lowest statistical
significance, but still it is greater than 3.1$\sigma$ and the
best-fit $f_{\rm H\emissiontype{I}}$ changes at most 8\%.
However this effect is quantitatively not negligible,
and one must be careful about the Ly$\alpha$ cross section
formulae in future studies. 

These results demonstrate that high precision Ly$\alpha$ damping wing
analyses of high-$z$ GRB afterglows are a powerful approach to measure
IGM H\emissiontype{I} fraction, which is sensitive to a small value of
$f_{\rm H\emissiontype{I}} \sim 0.05$.  The same results were obtained
by the two spectra taken by the two different telescopes, when the
model fitting method is the same, indicating that the systematic
uncertainty of the spectral measurement can be controlled to allow
discrimination of sub-percent level relative flux change in the
spectral shape.  However, a high precision fitting is easily affected
by systematic uncertainties about the fitting methods, especially the
wavelength range. Therefore one must be very careful to minimize the
systematic uncertainties, e.g. by checking the sensitivity of the
results to the adopted wavelength ranges and other model parameters.

\begin{ack}
We would like to thank O. Hartoog, J.P.U. Fynbo, P. Goldoni, and
T. Goto for providing us with their VLT spectrum and fruitful
discussions.  This work was supported by JSPS KAKENHI Grant Number
15K05018.
\end{ack}

\end{document}